# Absence of higher than 6-fold coordination in glassy GeO₂ up to 158 GPa revealed by X-ray absorption spectroscopy

**Authors:**

João Elias F. S. Rodrigues[1,‡], Angelika D. Rosa[1,§], Emin Mijit[1], Tetsuo Irifune[2], Gaston Garbarino[1], Olivier Mathon[1], Raffaella Torchio[1], Max Wilke[3]

**Affiliations:**

[1] European Synchrotron Radiation Facility (ESRF), 71 Avenue des Martyrs, 38000 Grenoble, France.

[2] Geodynamics Research Center, Ehime University Johoku Campus, 2-5 Bunkyocho, Matsuyama, Ehime 790–0826, Japan.

[3] Institute of Geosciences, University of Potsdam, Karl-Liebknecht-Strasse 24–25, 14476 Potsdam-Golm, Germany.

**Corresponding Authors:**

‡ rodrigues.joaoelias@esrf.fr

§ angelika.rosa@esrf.fr

**Significance**

The compression behavior of simple amorphous binary oxides, such as $GeO_2$ and $SiO_2$, above 1 Mbar remains highly controversial, despite its significance for understanding the compressibility of melts in Earth's deep interior and establishing the systematics of pressure-induced phase transitions in solids. Major challenges in obtaining robust evidence include the weak scattering signal of amorphous materials, the often-small sample volumes probed, and the subtle structural changes occurring at extreme pressures. X-ray absorption spectroscopy is highly sensitive to minor changes at the local atomic level, such as the variations in bond length, coordination number, and structural distortions. As such, it is an ideal technique for studying amorphous materials at extreme pressure conditions. In this study, we applied XAS to probe glassy $GeO_2$ beyond 1 Mbar pressure. The resulting data resolve previous controversies by providing clear criteria for distinguishing between the increase of polyhedral coordination, distortions, and connectivity in highly compressed glasses.

**Abstract**

Simple binary oxide glasses can exhibit a compression behavior distinct from that of their crystalline counterparts. In this study, we employed high-pressure X-ray absorption spectroscopy coupled to the diamond anvil cell to investigate in detail local structural changes around Ge in glassy $GeO_2$ up to 158 GPa. We conducted four independent runs, both with and without pressure-transmitting media. Up to 30 GPa, we observed no significant influence of the pressure medium on the pressure dependence of the Ge–O bond length ($<R_{Ge-O}>$). Between 10 and 30 GPa, the evolution of $<R_{Ge-O}>$ shows substantial variability across our experiments and previous works. The measured values lie close to those reported for crystalline polymorphs, including the rutile- and $CaCl_2$-type phase of $GeO_2$. This finding suggests that the amorphous structure possesses considerable flexibility to transition among different atomic configurations. From 30 GPa to 158 GPa, our results for both $<R_{Ge-O}>$ and the non-bonded cation-cation distance $<R_{Ge\cdots Ge}>$ demonstrate that edge-sharing octahedra remain the main structural motifs in glassy $GeO_2$. Up to 100 GPa, compaction proceeds primarily *via* distortions of octahedral O–Ge–O bond angles accompanied by octahedral bond shortening. Above 100 GPa, octahedral distortion becomes the prevailing mechanism. Compared to its crystalline analogues ($\alpha$-$PbO_2$ and pyrite-like phase), glassy $GeO_2$ exhibits a slightly less efficient compaction mechanism, likely due to kinetic constraints that inhibit reconstructive lattice rearrangements.

# 1. Introduction

The compression behavior of simple binary systems, such as $GeO_2$, $SiO_2$, $SnO_2$, $PbO_2$, $MgF_2$, and $CaCl_2$ has attracted much attention in the past 30 years (1-3). This interest arises for several reasons. First, a deep understanding of the pressure-induced progression of crystalline high-pressure phases enables the establishment of systematic behaviors, which can then be used to predict high-pressure structures in more complex systems or at ultra-high pressures. Notably, strong similarities in the high-pressure structural sequences have been observed across these binary systems. Second, understanding the densification mechanism of amorphous $SiO_2$ and the comparison with its crystalline counterpart has important implications for planetary evolution models (4-7). Specifically, the density difference between silicate melts and crystalline host rocks influences the mobility of deep melts with wide-ranging consequences for the geodynamic evolution of rocky planets (8-10). Third, improving our understanding of amorphous-amorphous transitions is crucial for designing and optimizing glasses for modern high-tech devices and for better understanding the phase diagrams of life-forming molecules, such as $H_2O$ (11), sulfur (12), and phosphorus (13). In these systems, a rich variety of molecular and polymeric liquid phases has been reported, potentially separated by sharp liquid-liquid transitions, though this remains subject of ongoing debate.

In the present work, we investigated the compression behavior of glassy $GeO_2$ (here referred to as $g$-$GeO_2$) and compare it in detail to observations of its crystalline counterpart. $GeO_2$ is a structural analogue of $SiO_2$ and its structural systematics are rather well-established at ambient and moderate pressure. Under ambient conditions, crystalline $GeO_2$ exists in two polymorphic phases: the quartz- and rutile-type structure. Similar to $SiO_2$, the quartz-type structure of $GeO_2$ consist of corner-sharing tetrahedral units, where each tetrahedron is linked to four others, forming a fully polymerized network. A structural transition to an intermediate monoclinic phase occurs above 6 GPa, in which Ge is fully in sixfold coordination (14-16). The rutile-type structure is only achieved at 22 GPa followed by a transition to the $CaCl_2$-type structure above 25 GPa, both transitions require laser-annealing (17, 18). The emergence of an intermediate fivefold coordination has not been observed upon compressing quartz-type $GeO_2$. Above 36 GPa, $GeO_2$ transforms into the $\alpha$-$PbO_2$-type phase showing highly distorted octahedral bonds around Ge, while above 65.5 GPa it forms the pyrite-type phase with a $(6 + 2)$ coordination Ge environment (19-23). These observations indicated that the sequence of high-pressure structures in crystalline $GeO_2$ follows the one found for $SiO_2$ and other binary systems, namely rutile-type $\rightarrow$ $CaCl_2$-type $\rightarrow$ $\alpha$-$PbO_2$-type $\rightarrow$ pyrite-type. However, all observed phase transitions in crystalline $GeO_2$ beyond the quartz-type structure required laser-annealing which often leads to only partial transformation. Therefore, only a few data points with precise crystallographic parameters such as atomic position have been reported that show high Ge–O bond length variations likely due to quenching effects. At pressures beyond 300 GPa, a post-pyrite phase of $GeO_2$ has been theoretically predicted in which Ge is ninefold coordinated (24).

In contrast to its crystalline counterpart, $g$-$GeO_2$ adopts only the quartz-like arrangement at ambient pressure (25-27). Upon increasing pressure and up to 40 GPa, the coordination number of Ge increases gradually from fourfold to eventually fivefold and sixfold in $g$-$GeO_2$ (4, 25, 26, 28-33). The occurrence of fivefold coordinated Ge and the completion of the transition to sixfold coordination remains, however, controversial. Literature data from experimental (4, 25, 26, 28-33) and computational work (34-36) on $g$-GeO$_2$ report significant dispersion of the average Ge–O bond length ($<R_{Ge-O}>$) at pressures above 10 GPa. These discrepancies were often attributed to differences in experimental techniques employed, which have different sensitivities for the local versus the bulk atomic structure. For example, the X-ray diffraction (XRD) study of Kono, *et al.* (4) reported a different pressure evolution of $<R_{Ge-O}>$ compared to studies employing X-ray emission (XES) (33) and X-ray absorption spectroscopy (XAS) (25, 28, 29, 31). XRD is primarily sensitive to the average bulk structure, while XAS and XES probe the local atomic structure. Notably, even among techniques with similar local sensitivities, such as XES and XAS, discrepancies persist, suggesting that the observed variations cannot be explained solely by differences in experimental methods.

Similar to the pressure evolution of $<R_{Ge-O}>$, there is presently no consensus regarding the evolution of the Ge coordination number in $g$-GeO$_2$ under compression above 30 GPa. Based on XRD data, Kono, *et al.* (4) proposed that $g$-GeO$_2$ adopts a higher than sixfold coordination already above 40 GPa and reaches a coordination of 7.2 at 93 GPa. This observation was corroborated by molecular dynamics simulations (MD) (34), that predicted a Ge coordination environment beyond sixfold above 20 GPa. Contrastingly, the persistence of sixfold coordinated Ge up to 100 GPa was suggested from XAS data (29, 31), valence-to-core XES measurements (33), and MD simulations (35, 36). This conclusion is also supported by high-pressure XAS data on complex aluminogermanate and germanium bearing silicate glasses up to 164 GPa (37, 38).

At present, only two experimental studies directly determined the density of $g$-SiO$_2$ and $g$-GeO$_2$ up to 133 GPa experimentally (10, 32). These studies concluded that the glassy network is less dense than its crystalline counterpart in the investigated pressure range. This result contrasts to previous experimental work that used indirect density determination methods and that reported a higher density of $g$-GeO$_2$ and $g$-SiO$_2$ compared to their crystalline counterparts (4, 5). These discrepancies may be related to the experimental difficulties to evidence structural changes in glasses at extreme pressures, because of their weak scattering signal, the small sample volumes probed in such experiments, and the subtle structural changes at extreme pressures. X-ray absorption spectroscopy is highly sensitive to fine changes at the local atomic scale, such as the average bond length and its variance (25, 27, 39). It constitutes, therefore, an ideal probe technique for investigating amorphous materials at extreme pressures. Here, we employed this method to probe glassy GeO$_2$ beyond Mbar pressures. The presented data are of very high quality (Fig. S1) and allowed resolving previous controversies as they provide clear criteria for the distinction between the increase of polyhedral coordination, distortions, and connectivity in highly compressed glasses.

## 2. Results

### 2.1. XANES observations

The energy region of the X-ray Absorption Near Edge Structure (XANES) comprises the rising edge, white line (WL), and absorption feature within few tens of eV above the WL. The XANES region includes information about the atomic organization due to significant intensity of the higher order (multiple) scattering terms of the low-energy photoelectrons. Therefore, changes of the electronic and atomic structural arrangement can be qualitatively identified from the evolution of XANES features. To this end, representative pressure-dependent normalized

Ge *K*-edge XANES spectra of *g*-GeO$_2$ are plotted in Fig. 1(a-d) for two runs conducted with and without a pressure transmitting medium, respectively (additional data are shown in Fig. S2 of the Supporting Information). At room temperature, the absorption edge is located between 11,110–11,112 eV, the WL at 11,110–11,117 eV, followed by two characteristic features $\alpha$ and $\beta$ in the energy ranges 11,118–11,126 eV and 11,128–11,138 eV, respectively. In all runs, we observe significant pressure-induced changes in the energy positions of the absorption edge, the WL, and the two features $\alpha$ and $\beta$. A previous XANES study on crystalline quartz-type GeO$_2$ associated the features $\alpha$ and $\beta$ as a signature of the [GeO$_4$] tetrahedral units (16). Beyond 10 GPa, $\alpha$ and $\beta$ disappear and a new broader feature here referred to as $\gamma$ emerges [Fig. 1 (a,b)]. Experimental and computational work have demonstrated that the $\gamma$ XANES feature is characteristic for 6-fold coordinated [GeO$_6$] units found in both the high-pressure rutile ($P4_2/mnm$) and metastable monoclinic ($P2_1/c$) phases of crystalline GeO$_2$ (16, 27).

To better characterize these pressure-induced changes, we determined the relative energy shift of the absorption edge [here called $\Delta E(P)$ to its ambient pressure value using $\Delta E(P) = E(P) - E(P_0)$, where $P_0 = 1$ bar. The edge position $E(P)$ was determined from the maximum of the first derivative of the XANES spectrum. The pressure evolution of $\Delta E(P)$ is shown in Fig. 1(e). It exhibits four discontinuities. Interestingly, overall changes of $\Delta E(P)$ are accompanied by changes in the WL edge energy position and width [Fig. 1(a-d)]. Up to 10 GPa, the absorption edge increases only slightly with 0.011 eV per GPa. In this interval, the features $\alpha$ and $\beta$ disappear. Between 10–30 GPa the absorption edge energy increases drastically by 0.042 eV per GPa and the $\gamma$ feature appears (panel b). Beyond 30 GPa and up to 100 GPa, the absorption edge energy increases more moderately with 0.016 eV per GPa. Above 1 Mbar and up 162 GPa the absorption edge energy does not change anymore significantly and only by 0.008 eV per GPa. A fifth region is highlighted as hatched light grey area and marks the pressure range in which X-ray-induced crystallization was observed (159–162 GPa, see Chapter 2.2 below).

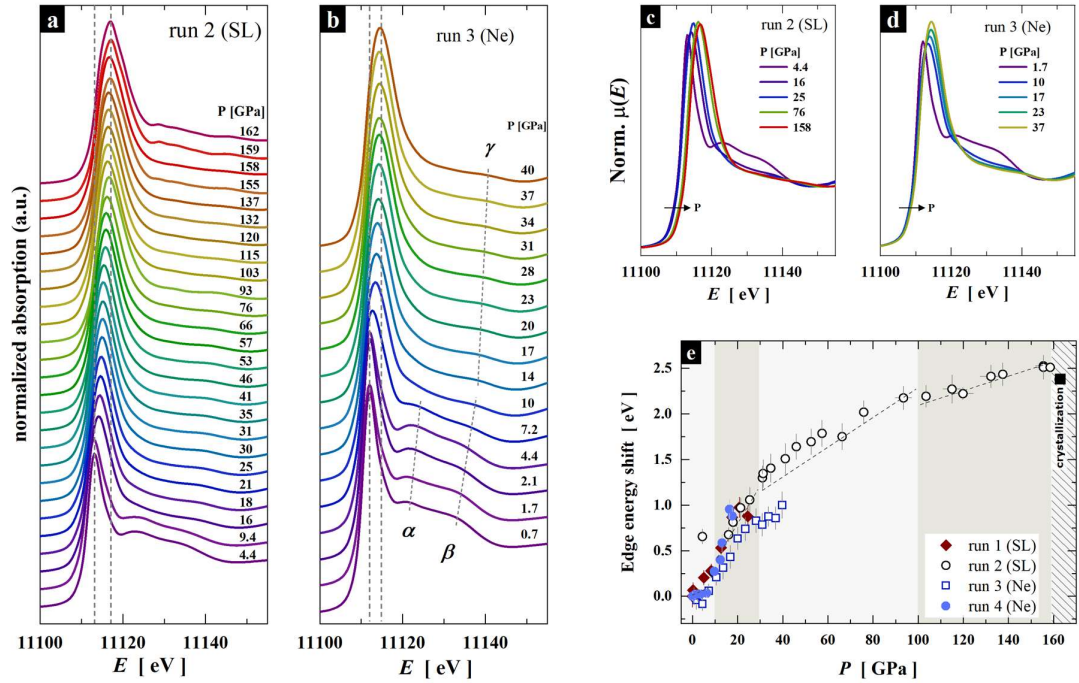

**Figure 1.** **Panels (a) and (b):** Normalized and vertically stacked Ge *K*-edge XANES spectra of run 2 (a) and run 3 (b) as a function of pressure (see Table 1 in the section Materials and Methods for more detailed information). In panel (a), the abbreviation SL indicates solid loading. The two vertical dashed lines delineate the white line positions at the lowest and highest probed pressure, respectively. The Greek letters $\alpha$, $\beta$, and $\gamma$ above the spectra indicate distinct spectral features. Dashed lines serve as guides for the eye to follow the features' evolution with pressure. **Panels (c) and (d):** Selected high-pressure XANES spectra plotted without vertical offset for run 2 and run 3. **Panel (e):** Calculated edge energy shift $\Delta E(P)$ for all runs. Dashed gray lines are guides for the eye. Four different pressure intervals can be distinguished using discontinuous changes seen in the pressure evolution of $\Delta E(P)$. These regions are delimited here by light and dark grey background colors.

## 2.2. EXAFS observations

The Extended X-ray Absorption Fine Structure (EXAFS) part of the absorption spectrum follows the XANES region at higher energy. It exhibits characteristic oscillations caused by the scattering of the photoelectron emitted by the absorbing atom on neighboring atoms. The signal can be decomposed in sine waves using the Fourier transformation ($kR$ spaces), each corresponding to one scattering phenomena. The frequencies of these sine waves are related to the interatomic distances, while the amplitude is related to the sort and number of neighboring atoms in different coordination shells, as well as to the structural disorder or bond length variance. If the structural disorder is high, as it is the case for glasses, the signals' amplitude is reduced significantly. Therefore, a very high signal-to-noise ratio is required to extract these weak signals. This presents a technical challenge for high-pressure XAS studies, as it demands a highly focused beam, a high beam photon flux, and a very high beam position stability during the energy scan. These requirements are met at BM23 (X-ray Absorption Beamline) of the ESRF-EBS (40, 41).

In Fig. 2 and Fig. S3, representative raw $k^3$-weighted EXAFS functions [Fig. 2(a)] and their respective Fourier transforms (here referred to as *FT*) [Fig. 2(b)] are plotted. EXAFS functions below ~10 GPa are characterized by regular and symmetric oscillations, resulting in a single feature at ~1.25 Å in their *FTs*. This feature can be associated with the next nearest neighboring atoms to Ge and thus to oxygen. Qualitatively, the peak position in the *FT* provides information on the average Ge–O bond distance $<R_{Ge-O}>$, its width and amplitude information on the distance variance ($\sigma^2_{R_{Ge-O}}$) and the number of coordinating neighbors ($N_{Ge-O}$). Above 10 GPa, the EXAFS oscillation between 6 and 8 Å$^{-1}$ splits into two peaks and a second but weaker oscillation appears at *k*-values above 10 Å$^{-1}$ [Fig. 2(a)]. This splitting results in the appearance of an additional weak signal in the *FT* at higher *R* values around 2.5 Å [Fig. 2(b)]. This weak signal persists in the *FT* up to the highest probed pressure and can be associated with the second nearest neighbor of the absorbing Ge atom and thus to the average distance between two Ge atoms $<R_{Ge\cdots Ge}>$. EXAFS data recorded at 162 GPa are presented in Fig. S4 and exhibit clear changes, including sharpening of all the oscillations and the emergence of additional oscillations. We associated these new features with X-ray beam-induced crystallization. This interpretation was confirmed through X-ray diffraction (XRD) measurements (see Fig. S5).

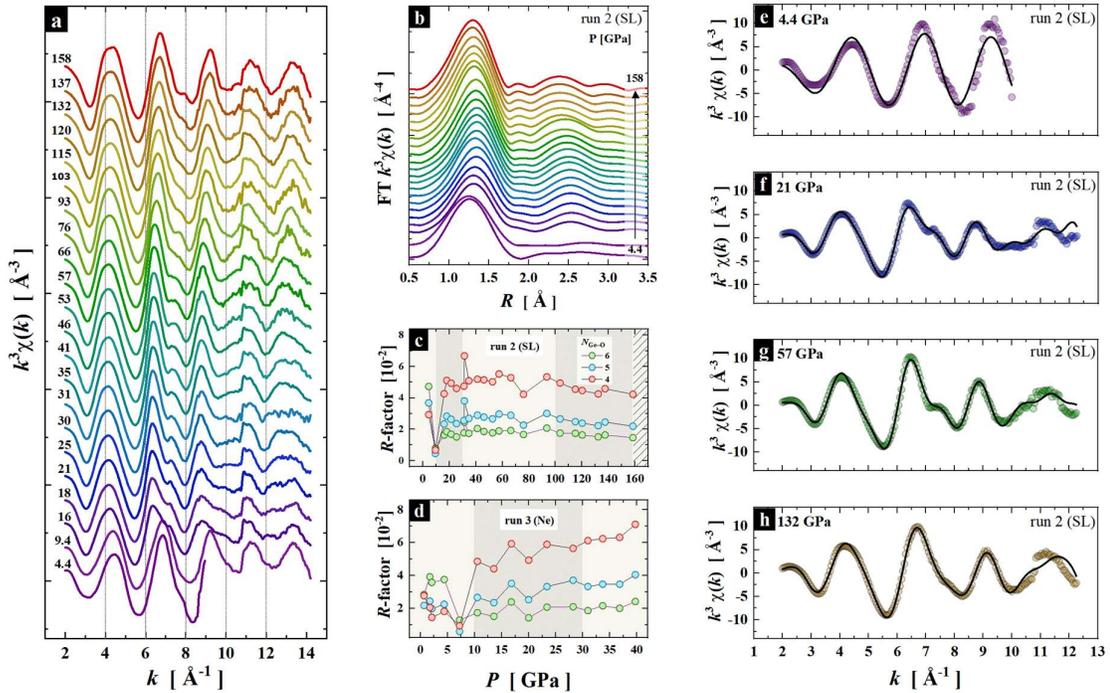

**Figure 2.** Panels (a) and (b): Pressure evolution of raw $k^3$-weighted EXAFS oscillations $\chi(k)$ of run 2 and their *FT* magnitudes (not phase-shift corrected). Panels (c) and (d): *R*-factors from best fittings considering fixed ($N_{Ge-O}$) of 4, 5, and 6 for runs 2–3, respectively. The background colors follow the division of pressure intervals as in Fig. 1. Panel (e): Comparison between experimental and best fit data of run 2 at selected pressure points. Experimental points are plotted as open symbols, while black solid lines represent the best EXAFS fit considering ($N_{Ge-O}$) as free parameter (fitting method 1, see text and Supporting Information).

To gain quantitative information on the pressure evolution of the local and medium range atomic structure in *g*-GeO$_2$, we performed detailed EXAFS fittings to all data. To this end, we conducted for each spectrum two fits with different structural input models. The first model comprised only one scattering path between Ge and O and fitted parameters $<R_{Ge-O}>$, $\sigma^2_{R_{Ge-O}}$, and $N_{Ge-O}$. The second model comprised two additional scattering paths: one between Ge and Ge atoms in the 2$^{nd}$ coordination shell and one between Ge and O atoms in the 3$^{rd}$ coordination shell. The second model includes additional 6 fit parameters: $<R_{Ge\cdots Ge}>$, $\sigma^2_{R_{Ge\cdots Ge}}$, $N_{Ge\cdots Ge}$, $<R_{Ge\cdots O}>$, $\sigma^2_{R_{Ge\cdots O}}$, and $N_{Ge\cdots O}$; however, $N_{Ge\cdots Ge}$ and $N_{Ge\cdots O}$ were kept as fixed to 2 and 4, respectively. We further performed a careful evaluation of potential correlations between fitted parameters (see Supporting Information for more details). These tests revealed that the parameter ($N_{Ge-O}$) can be set as a free parameter, but that uncertainties on absolute values of this parameter are still high due its high correlations to other fit parameters [see Fig. 2(c-d)]. The resulting fit values for all parameters and for all runs are listed Table S1. In Fig. 3(a), a comparison between the raw $FT$ and the calculated $FT$ from the best-fitted structural model is shown at selected pressures. As can be seen in Fig. 3(a), the fitted structural model describes very well the raw data up to 158 GPa.

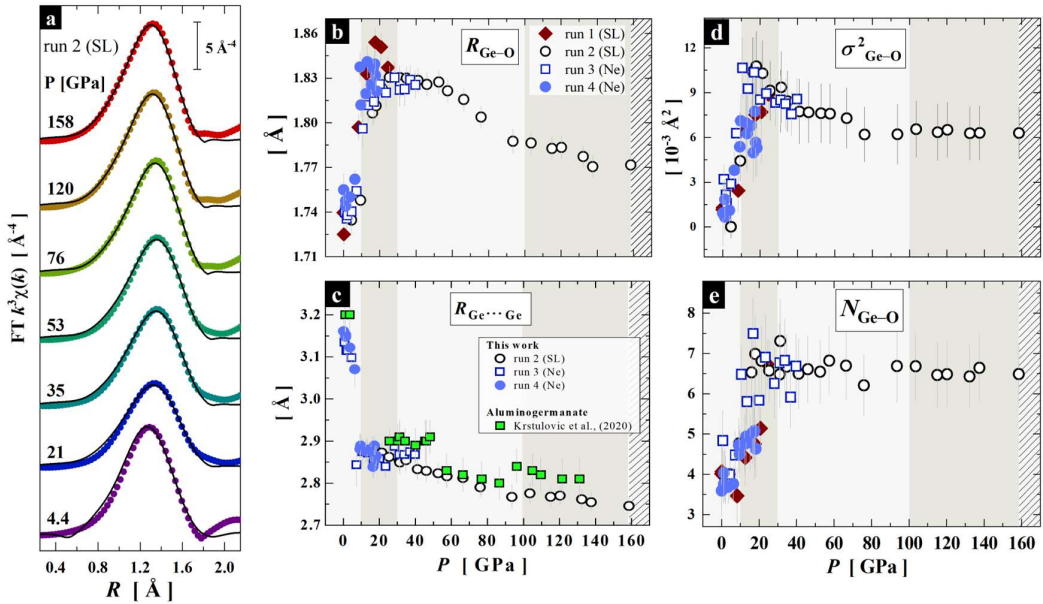

**Figure 3.** **Panel (a):** Evolution of the raw $FT$ magnitudes (circular symbols) with pressure and the corresponding best EXAFS fits (solid lines) for run 2. **Panels (b) and (c):** Pressure dependence of the average first neighbor bond distance $<R_{Ge-O}>$ and the non-bonded cation-cation distance $<R_{Ge\cdots Ge}>$, the latter is compared to reported data of aluminogermanate glasses by Krstulovic, *et al.* (37). **Panels (d) and (e):** Pressure evolution of the Ge–O bond length variance $\sigma^2_{R_{Ge-O}}$, and the Ge coordination number $N_{Ge-O}$. The background colors in panels (b-e) follow the division of pressure intervals as in Fig. 1.

In Fig. 3(b-e), the pressure evolutions of fitted structural parameters including $<R_{Ge-O}>$, $<R_{Ge\cdots Ge}>$, $\sigma^2_{R_{Ge-O}}$, and $N_{Ge-O}$ are presented. In all the runs, $<R_{Ge-O}>$ expands first from ambient pressure to 30 GPa from ~1.72 to 1.83 Å ($\Delta R$ ~ 0.1 Å) [panel 3(c)]. Above 30 GPa, $<R_{Ge-O}>$ progressively shrinks up to 100 GPa, before it stagnates up to 158 GPa. Both $\sigma^2_{R_{Ge-O}}$ and $N_{Ge-O}$ follow a similar behavior as $<R_{Ge-O}>$ up to 30 GPa. They undergo an increase of their ambient condition value from ~0.002(7) to 0.008(7) Å$^2$ and from 4.0(7) to 6.6(7), respectively. Beyond 30 GPa, $\sigma^2_{R_{Ge-O}}$ and $N_{Ge-O}$ remain rather constant. The non-bonded Ge distance $<R_{Ge\cdots Ge}>$ remains constant around 3.10(2) Å from ambient pressure to 6 GPa. Beyond this pressure, it drops suddenly to ~2.88 Å and progressively decreases further to a value of ~2.77 Å around 100 GPa. Between 100 and 158 GPa, $<R_{Ge\cdots Ge}>$ remains invariant. We noticed, that the discontinuities seen in structural parameters follow those qualitatively observed for the absorption edge and WL energy position (see Fig. 1). This indicates a strong consistency among both qualitative and quantitative analysis results obtained from XANES and EXAFS, respectively. In the investigated pressure range, no clear effect of PTM can be observed in the pressure evolution of fitted parameters for different runs, this result is in line with those of previous works on glassy $GeSe_2$ (42, 43).

## 3. Discussion

The discontinuities observed in the pressure evolution of XANES features and fitted EXAFS variables (Fig. 1–3, Table S1) clearly define four distinct pressure regimes of the compression behavior of g-$GeO_2$, namely: (*i*) 0–10 GPa, (*ii*) 10–30 GPa, (*iii*) 30–100 GPa, and (*iv*) 100–158 GPa. In the following paragraphs, we compare our observations for each interval with prior experimental (XAS, XES, XRD), theoretical, and crystalline $GeO_2$ data. To infer the underlying structural rearrangement mechanisms in the $GeO_2$ glass, we relate our observations to established theories that describe structural transitions in solids and glasses, including the bond valence theory of Brown and Altermatt (44), Brown (45), and the cation-cation repulsion theory of O'Keeffe and Hyde (46). More details on these theories can be found in the Support Information. We also would like to point out that the average bond distance $<R_{Ge-O}>$ is the most reliable structural parameter extracted from EXAFS, followed by the parameters $<R_{Ge\cdots Ge}>$ and $\sigma^2_{R_{Ge-O}}$. In contrast, the pressure evolution of $N_{Ge-O}$ can only be used for a qualitative comparison, as the absolute value is highly correlated and affected by high uncertainties of 10–15%.

### 3.1 First compression stage between 0 and 10 GPa

In the first compression stage up to 10 GPa, $<R_{Ge-O}>$, $N_{Ge-O}$, and $\sigma^2_{R_{Ge-O}}$ progressively increase while $<R_{Ge\cdots Ge}>$ gradually reduces (Fig. 3). The observed variation in $<R_{Ge-O}>$ can be qualitatively linked to the coordination number variations using the bond valence theory. After Brown and Altermatt (44), the bond valence is calculated from the cation-anion distance by:

$$s = \exp\left(\frac{R_0 - R}{B}\right) \tag{1}$$

where $R_0$ and $B$ are the fitted bond valence parameters describing the nominal length of a bond of unit valence and the softness of interaction of two atoms, respectively. After Pauling's second rule, the sum of bond valences around a given cation or anion should be equal to the formal charge. Tetrahedrally coordinated Ge in g-$GeO_2$ fulfills this rule but represents a virtually incompressible unit. A further reduction of $<R_{Ge-O}>$ would lead to over bonding of the bridging oxygens, resulting in an unrealistic bond valence sum around oxygen greater than 2. Thus, during the initial compression stage structural densification proceeds via rearrangement of [$GeO_4$] units, up to the limit imposed by Ge–Ge repulsive forces. At higher pressures, further compaction is achieved by shifting the balance between non-bonded repulsive forces and attractive forces via the creation of additional cation anion bonds [see O'Keeffe and Hyde (46)]. Therefore, corner-shared [$GeO_4$] tetrahedra with oxygen CN = 2 transform into edge-shared octahedra where oxygen has a CN of 3.

In the context of these considerations, the increase of $<R_{Ge-O}>$ suggests the formation of higher coordinated cation polyhedra, and the reduction of $<R_{Ge\cdots Ge}>$ is consistent with the gradual increase of the anion coordination. Our data confirm therefore previous observations that in $g$-GeO$_2$ corner-sharing [GeO$_4$] tetrahedra start rotating against each other to form corner- and edge-sharing [GeO$_5$] and [GeO$_6$] polyhedra (26, 47). Pressure facilitates the gradual reduction of distance between non-bonded and normally highly repulsive Ge cations (Ge$\cdots$Ge), while the increased number of bonds stabilizes the new structural arrangement. In crystalline GeO$_2$, this transformation occurs above 6 GPa at ambient temperature, driven by tetrahedral rotations involving bond breaking and cooperative atomic shuffling, consistent with a martensitic-like transition mechanism (15).

### 3.2 Second compression stage between 10 and 30 GPa

Above 10 GPa, $<R_{Ge-O}>$ increases significantly and reaches a maximal value of 1.851(10) Å around 20 GPa in run 1. In runs 2–4, a maximum elongation of ~1.830(8) Å is reached around 30 GPa. These observations indicate that, above 10–30 GPa, $g$-GeO$_2$ transitions further to sixfold coordinated Ge. The observed increase of $<R_{Ge-O}>$ in all runs between 10–20 GPa could suggest also the formation of fivefold [GeO$_5$] coordinated units that coexist with fourfold and sixfold coordinated Ge. The present XANES and EXAFS data alone can, however, not unambiguously prove or disprove the presence of fivefold coordinated Ge. Indeed, only advanced numerical simulations that consider the live-time of a given coordination in the MD run could provide deeper insights (48), which are out of the scope of the present study. Previous numerical studies did not consider this parameter (34-36). They report that most of [GeO$_4$] units and, if present [GeO$_5$] polyhedral have already evolved to an octahedral coordination around 10–15 GPa. The present observations from EXAFS are supported by those from XANES. At 10 GPa a broad feature ($\gamma$) appears [Fig. 1(c)], which has been associated previously to sixfold coordinated Ge while concomitantly the features $\alpha$ and $\beta$ disappear that are representative of 4-fold coordinated Ge (16, 27). Interestingly, $<R_{Ge-O}>$ distances decrease beyond 20 GPa only for run 1 and reach a value close to those of runs 2–4 at 25 GPa, which is the highest probed pressure for run 1. Beyond 30 GPa, $<R_{Ge-O}>$ distances decrease progressively also for runs 2–4, which marks therefore the end of the second structural rearrangement stage.

Fitted $<R_{Ge-O}>$ distances in $g$-GeO$_2$ exhibit a high scatter in between runs above 10 to 30 GPa. This variability may reflect substantial structural disorder and or the presence of non-Gaussian bond-length distributions. However, EXAFS fits that included non-Gaussian distributions did not improve the fit quality. While incorporating this parameter, systematically increased distances were obtained that still could not account for the large scatter. Likewise, only minor variations were found in the fitted $E_0$ parameter during the EXAFS analysis (see Table S1), leading us to conclude that this factor also does not explain the dispersion of $<R_{Ge-O}>$ values in this pressure range. A potential explanation for the observed dispersion could be the high bond-length variability of [GeO$_6$] octahedra, as reported above 10 GPa in rutile-type GeO$_2$ (14) [see small yellow symbols in Fig. 4(a)].

To better understand the observed pressure evolutions of $<R_{Ge-O}>$ from EXAFS, we compared it to those reported for crystalline GeO$_2$ polymorphs [Fig. 4(a)]. In fact, $<R_{Ge-O}>$ distances in $g$-GeO$_2$ approach closely those reported for crystalline counterparts comprised of sixfold coordinated Ge octahedra between 10 and 30 GPa, including the monoclinic-type (edge-sharing octahedral chains, with partial occupations), the rutile-type and the $CaCl_2$-type arrangement (both edge- and corner-sharing octahedral, with 2 short axial and 4 long equatorial bonds) (14, 18, 20). In crystalline $GeO_2$, the $CaCl_2$-type phase evolves gradually from the rutile-type phase through a second-order displacive transition between 25 and 27 GPa. The structural densification is achieved through octahedra rotation, which results in a rearrangement of the oxygen sublattice from a distorted to highly ordered hexagonal close packed arrangement. Above 16 GPa, octahedral rotation is coupled with octahedral distortion, indicated by the greater variance in bond angles and bond lengths [small yellow and red symbols Fig. 4(a)]. Because the $<R_{Ge-O}>$ values in $g$-$GeO_2$ closely approach those reported for crystalline $GeO_2$ between 10 to 30 GPa, we propose that, on a local scale, $g$-$GeO_2$ may gradually evolve through octahedral arrangements resembling those of its crystalline counter parts.

The observed high scatter in $<R_{Ge-O}>$ values between runs could also indicate that g-$GeO_2$ may follow different structural pathways during the displacive transition from the rutile- to $CaCl_2$-type structures. To examine this hypothesis, we compare the pressure evolution of $<R_{Ge-O}>$ values for runs1–4 with those reported in previous experimental studies employing various techniques in Fig. 4(b). Interestingly, reported pressure trends vary but generally fall within the scatter observed across our runs 1–4. Previously, such differences were attributed mainly to variations in experimental probes. However, we found strong consistency for the pressure evolution of $<R_{Ge-O}>$ values of run 1 to a previous XES (33), XRD (4) study and two molecular dynamic simulations (35, 36). XES - a technique similar to XAS and insensitive to anharmonicity or non-Gaussian effects, while XRD probes bulk structure differently from XAS and XES - closely matches our run 1. Interestingly, more recent XAS studies, including those by Baldini, *et al.* (29) and Hong, Newville, Duffy, Sutton and Rivers (31), show pressure trends resembling either run 1 or runs 2–4. The absolute $<R_{Ge-O}>$ values in these works are slightly lower than ours, likely due to differences in EXAFS fitting strategies and signal length. From this comparison, we conclude that differences in $<R_{Ge-O}>$ evolution between 10 and 30 GPa cannot be explained solely by methodological factors. Instead, they are likely to reflect the high structural flexibility of glassy $GeO_2$, which may transition through multiple pathways towards the $CaCl_2$-type arrangement. Further exploration of this hypothesis will require detailed experimental and computational studies that integrate multiple probe techniques.

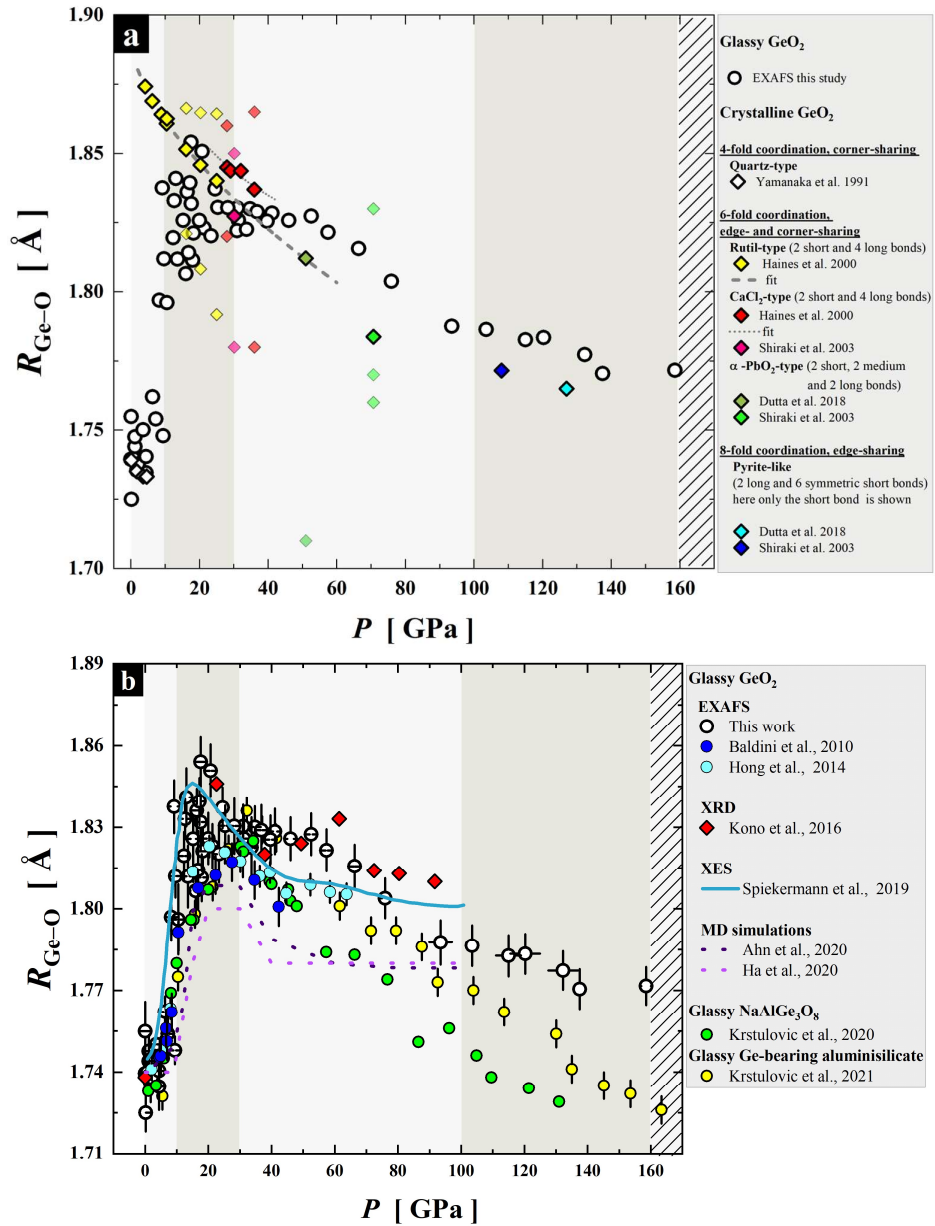

**Figure 4. Panel (a):** The pressure evolution of $<R_{Ge-O}>$ (gray circles) extracted from EXAFS for *g*-GeO$_2$ to those reported for crystalline GeO$_2$ polymorphs from XRD (colored diamond symbols: quartz-, rutile-, CaCl$_2$-, α-PbO$_2$-, and pyrite-type phases). For clarity, the uncertainties on $<R_{Ge-O}>$ are not shown (but can be found in panel (b)). Note, that for XRD studies the average bond length is shown as big diamond symbols, while small pale diamond symbols of the same color represent individual bond length for a polyhedron. **Panel (b):** Comparison of $<R_{Ge-O}>$ values reported for *g*-GeO$_2$ from this work, from more recent EXAFS studies published after 2010, including Baldini, *et al.* (29) and Hong, Newville, Duffy, Sutton and Rivers (31), from the XRD study of Kono, *et al.* (4), from the XES study of Spiekermann, *et al.* (33), and from molecular dynamics (MD) simulations (34, 36). The Ge–O bond distances derived from aluminogermanate and aluminosilicate glasses are also shown for comparison (37, 38). Uncertainties are within symbol sizes if not shown.

## 3.3 Third compression stage between 30 and 100 GPa

The third pressure interval between 30 and 100 GPa is characterized by a gradual decrease of the parameters $<R_{Ge-O}>$, $\sigma^2_{R_{Ge-O}}$, and $<R_{Ge\cdots Ge}>$. *Ab-initio* calculations predicted a structural phase transition at 36 GPa from the CaCl$_2$-type to the $\alpha$-PbO$_2$-type in crystalline GeO$_2$ (19). In the $\alpha$-PbO$_2$-type phase, the anions remain in hexagonal-closed packed arrangement like the CaCl$_2$-type structure, but the Ge cations occupy half of the octahedral vacancies in an ordered way. The octahedra are highly deformed comprising 2 short and 2 medium Ge–O bonds in the equatorial plane and 2 long bonds in axial direction. In contrast to the rutile- and CaCl$_2$-type phase, the octahedra are not connected through the equatorial edges formed of long Ge–O bonds, but through an edge formed by a medium and long distant oxygen atom of the equatorial plane and the apex of the octahedra, respectively. This results in undulating "zick-zack" chains of highly deformed edge-sharing [GeO$_6$] octahedra. The octahedral chains are interconnected via corners. The transformation from the CaCl$_2$-type to the $\alpha$-PbO$_2$ phase requires strong structural rearrangements and notably diffusion of Ge cations and thus breakage of bonds.

Experimental studies have found evidence for this transformation above 41 GPa after laser-annealing of the sample at high temperatures ($T > 1200$ K), which leads only to a partial transformation of the sample (22, 23). The single-crystal XRD study of Dutta, White, Greenberg, Prakapenka and Duffy (23) reported interatomic distances for the $\alpha$-PbO$_2$ at 50.8 GPa, including an average $<R_{Ge-O}>$ distance of 1.81(10) Å and $<R_{Ge\cdots Ge}>$ distances ranging from 2.860(5) for edge-sharing to 3.187(2)Å for corner-sharing Ge polyhedral. At 70.7 GPa, Shiraki, Tsuchiya and Ono (20) found an average value for $<R_{Ge-O}>$ of 1.79(2) Å and $<R_{Ge\cdots Ge}>$ distances of 2.841(5)–3.147(2)Å between edge- and corner-sharing polyhedra.

Compared to results from XRD studies, we found slightly larger values for $<R_{Ge-O}>$ and slightly smaller values for $<R_{Ge\cdots Ge}>$, that agree however within their experimental uncertainties [Fig. 3(b) and Fig. 4(a)]. The distance reduction of $<R_{Ge-O}>$ above 30 GPa observed in run 1 is conform with results from XRD on $g$-GeO$_2$ up to 80 GPa (4) and with MD simulations up to 50 GPa [see Fig. 4(b)] (35, 36). Interestingly, partially polymerized germanate glasses exhibit similar values for $<R_{Ge-O}>$ and $<R_{Ge\cdots Ge}>$ compared to XRD studies (37), while reported $<R_{Ge-O}>$ values from Ge-bearing silicate glasses (38), in which Ge is a minor element, present significantly smaller values for $<R_{Ge-O}>$ compared to XRD data [Fig. 4(b)]. The higher compressibility of $<R_{Ge-O}>$ in the partially polymerized glasses can be explained with the presence of network-modifying elements, namely Na, Mg, Sr, or other elements that influence non-bonded forces between network-forming cations such as Al. The presence of these elements induces non-bridging oxygen atoms that break the interconnectivity of Ge polyhedra and, thus, may contribute to greater structural flexibility.

Overall, the short $<R_{Ge\cdots Ge}>$ distances below 3 Å, observed form EXAFS for $g$-GeO$_2$, can only be explained by the occurrence of edge-sharing octahedra. Any change of Ge coordination would imply an increase of $<R_{Ge\cdots Ge}>$ bond length according to bond valence rules (45). This fact suggests that this polyhedral connectivity remains dominant in $g$-GeO$_2$ up to the highest probed pressure (158 GPa). This conclusion holds also for aluminogermanate glasses [Fig. 3(b)]. The reduction of monotonous bond length variance ($\sigma^2_{RGe-O}$), observed for $g$-GeO$_2$ in the pressure interval 30–100 GPa, indicates a minimization of configurational entropy and an overall bond length symmetrization. This observation contrasts with the highly deformed Ge octahedra reported for the $\alpha$-PbO$_2$-type phase of crystalline GeO$_2$ stabilized in this pressure range. We propose that the stiffness between fully connected octahedral units in $g$-GeO$_2$ glass inhibits the reorganization of the Ge cations to the $\alpha$-PbO$_2$ structure-type. Therefore, $g$-GeO$_2$ may stabilizes rather in a CaCl$_2$-type arrangement upon compression. Overall, our results indicate that densification in $g$-GeO$_2$ involves shorter and more uniform Ge–O bonds, while the octahedral angles (O–Ge–O and Ge–O–Ge) distort significantly from their ideal values, causing octahedral rotation and bond angle changes. The latter mechanism can explain the observed decrease of the $<R_{Ge\cdots Ge}>$ interatomic distances.

We note, that this compression mechanism is less effective compared to the transformation to the $\alpha$-PbO$_2$-type phase because of the slightly longer average $<R_{Ge-O}>$ bond length observed for $g$-GeO$_2$ compared to its crystalline counterpart [Fig. 4(a)] (*i.e.*, at 70 GPa $<R_{Ge-O}>_{\text{EXAFS}}$ = 1.810(7) and $<R_{Ge-O}>_{Shiraki\_XRD}$= 1.79(2) Å). In contrast to $g$-GeO$_2$, $\sigma^2_{R_{Ge-O}}$ of aluminogermanate glasses stagnates at high values above 30 GPa and does not decrease with increasing pressure even up to 160 GPa. These observations may indicate the presence of highly distorted Ge octahedra in aluminogermanate glasses, exhibiting potentially a high variability of the bond length close to those observed in the $\alpha$-PbO$_2$-type phase of crystalline GeO$_2$. This and the similarities of other structural parameters compared to those reported from XRD of the $\alpha$-PbO$_2$ phase, including $<R_{Ge-O}>$ and $<R_{Ge\cdots Ge}>$, suggest that polymerized glasses (18, 19) reorganized into a more compact arrangement of octahedra above 30 GPa, most likely facilitated by the higher structural flexibility.

Kono, *et al.* (4) concluded from XRD data on $g$-GeO$_2$, that at 93 GPa Ge has reached a coordination number of 7.2. However, the coordination number extracted from XRD data is based on integrating the experimental pair distribution function $g(r)$ across a certain radius range. The result of this method is highly dependent on the chosen cutoff radius and is therefore also highly affected by peak overlap with higher coordination shells. Further, the reported increase of coordination number is inconsistent with their reported decrease of $<R_{Ge-O}>$ distances, which is also observed in the present EXAFS study up to 100 GPa, and which rules out any further increase of polyhedral coordination according to the bond valence theory (45).

Our conclusion differs from that of previous MD simulations. Du and Tse (34) suggested the competition of 6- and 7-fold coordinated units in the pressure interval, driven by the oxygen packing fraction. Ha and Kien (35) proposed the formation of [GeO$_6$] domains due to the progressive reduction of [GeO$_5$] units, without the occurrence of higher than sixfold coordinated polyhedral. Both mechanisms would require an increase or stagnation of $<R_{Ge-O}>$ with pressure, which is not observed in the present EXAFS data. Anh, Trang, Nguyet, Linh and Hong (36) proposes that the densification of $g$-GeO$_2$ is driven by the formation of edge- and face-sharing bonds between Ge octahedra, with relative fractions of 71% corner-sharing, 23% edge-sharing, and 5% face-sharing bonds per Ge atom. This mechanism could be plausible considering the observed continuous decrease of $<R_{Ge\cdots Ge}>$ to 100 GPa. However, Anh, Trang, Nguyet, Linh and Hong (36) also predicted very short $<R_{Ge-O}>$ bond distances for face- and corner-sharing Ge polyhedral of 2.2 and 2.5 Å, which are not noticed here. The formation of face-sharing units would be further in conflict with bond valance sum rules. Besides, in the rutile- and CaCl$_2$-type phases of crystalline GeO$_2$, one Ge atom is coordinated by 66.6% edge-sharing bonds and 33.3 % corner-sharing bonds. The extracted EXAFS data for $g$-GeO$_2$ fit very well with bond distances observed for these structure types up to 30 GPa [Fig. 4(a)], as discussed in the previous chapter. We therefore consider, that in $g$-GeO$_2$ the edge-sharing bond is predominant between 30 and 100 GPa. Discrepancies to numerical studies may be related to the strong bias in the assignment of coordination numbers if the live-time of a given coordination in the MD run is not considered (48).

### 3.4 Forth compression stage between 100 and 158 GPa

At pressures above 100 GPa and up to 158 GPa, only very slight reductions of $<R_{Ge-O}>$ distances and $<R_{Ge\cdots Ge}>$ distances are observed for g-GeO$_2$, while the values of $\sigma^2_{RGe-O}$ and $N_{Ge-O}$ are stagnating (Fig. 3). This behavior could suggest the formation of a very small fraction of higher than sixfold coordinated Ge polyhedral. Indeed, bond valence theory suggests that higher coordination results in a longer average bond length $<R_{Ge-O}>$ and greater bond variance $\sigma^2_{RGe-O}$. *Ab initio* calculations predicted the stabilization of the pyrite-type phase ($Pa\overline{3}$) in crystalline GeO$_2$ above 70 GPa (19). This phase has been experimentally observed in crystalline GeO$_2$ at 108 GPa. In the pyrite-type phase, Ge adopts a local coordination number of (6 + 2) (deformed cube or bicapped octahedron), including 6 oxygen atoms in a distance of 1.771(7) Å and two oxygen atoms in a distance of 2.567(7) Å, leading to a weighted average bond-distances of 1.97 Å (25). The average $<R_{Ge-O}>$ bond distance extracted from EXAFS in g-GeO$_2$ at 108 GPa equals ~1.785 Å. Considering that the EXAFS analysis provides the average bond distance, our results suggest that the possible presence of (6 + 2) coordination can be considered negligible. In addition, we consider that if a non-negligible amount of Ge in (6 + 2) coordination would be present in the sample, the respective EXAFS data could not be fitted using the present structural model that ignores the extra 2 oxygen atoms. This comparison indicates only a minor change in the structural arrangement in g-GeO$_2$ above 1 Mbar.

We further argue that the prevalence of short $<R_{Ge\cdots Ge}>$ distances suggests that edge-sharing octahedra are still the main structural motif in glassy GeO$_2$ above 100 GPa [Fig. 3(b)]. As stated above, the rutile-, CaCl$_2$-, and $\alpha$-PbO$_2$-type phases exhibit two $<R_{Ge\cdots Ge}>$ distances around ~2.86 Å and ~3.20 Å corresponding to distances of Ge atoms linked over edge- and corner-sharing octahedral, respectively. In the pyrite-like phase, Ge atoms form a face-centered cubic sublattice and exhibit therefore only one very regular $<R_{Ge\cdots Ge}>$ distance of 3.0664(11) Å between purely edge-sharing bicapped octahedra. If the pyrite-type structural elements stabilize, we would expect a shift of the two $<R_{Ge\cdots Ge}>$ distances towards 3.07 Å above 100 GPa, which is not observed in our data [Fig. 3(b)]. We exclude, therefore, that this structural motif is adapted in g-GeO$_2$. Based on the present EXAFS data, we propose an alternative compression mechanism above 100 GPa involving rotations of sixfold coordinated [GeO$_6$] units accompanied by significant bond length distortions in octahedral, similar to those observed for the $\alpha$-PbO$_2$ phase. This mechanism may become favorable above 100 GPa because of the significant over pressurization of the structure. The slightly higher $<R_{Ge-O}>$ observed for g-GeO$_2$ compared to its crystalline counterpart above 100 GPa, suggest again that the compaction mechanism of g-GeO$_2$ is less efficient. This is conclusion is conform with the observed lower density of g-GeO$_2$ above 100 GPa compared to crystalline phases (32).

Upon compression above 158 GPa, g-GeO$_2$ crystallized in the X-ray beam and formed the pyrite-like structure (Fig. S5). This observation attests the high strain and metastability of the glass structure compressed at ambient temperature to such high pressures. During the X-ray absorption measurements, Ge atoms absorb a substantial portion of the X-ray beam and are promoted to an excited electronic state. Together with the excess strain energy stored in the structure, this phenomenon supplies the metastable glass the energy needed to crystallize into a stable crystalline phase. Previous first principle calculations have predicted the stabilization of a cotunnite $\alpha$-PbCl$_2$-type structure of GeO$_2$ above 300 GPa in which Ge is 9-fold coordinated (24). Our experimental results show that the pyrite-like structure remains stable in crystalline GeO$_2$ up to 158 GPa. Further work is required to understand how $g$-GeO$_2$ behaves beyond 160 GPa and at high temperature.

## 4. Conclusions

In this work, we investigated the pressure-evolution of the local and medium range atomic structure in GeO$_2$ glass using X-ray absorption spectroscopy to 158 GPa. Between 10 and 30 GPa, our data suggest that glassy GeO$_2$ can transition via different pathways from the quartz- to the monoclinic-type, rutile-type, and CaCl$_2$-type structural arrangements. Such a mechanism could explain the observed discrepancies between previous works. Our data further demonstrate the predominance of edge-sharing octahedral environments in glassy GeO$_2$ up to 158 GPa. Above 30 and until 100 GPa, these Ge–O octahedra shorten and symmetrize, while octahedral bond angles (O–Ge–O and Ge–O–Ge) distort from ideal values. These observations contrast with reported highly distorted octahedral bonds reported for its crystalline counterpart, the $\alpha$-PbO$_2$-phase of GeO$_2$, which stabilizes above 41 GPa and after laser heating. Above 100 GPa, our data suggest that octahedral distortion becomes the predominant compaction mechanism, both in terms of bond length and octahedral bond angles. Once again, this observation contrasts with findings on crystalline GeO$_2$. Above 100 GPa, the pyrite-like structure has been reported, comprising bicapped octahedra characterized by six highly symmetric bonds and two significantly longer ones. These discrepancies can be explained by the reconstructive nature of the structural phase transitions from CaCl$_2$ $\rightarrow$ $\alpha$-PbO$_2$ $\rightarrow$ pyrite-like phase in crystalline GeO$_2$. These reconstructive transformations are inhibited in the highly polymerized cold compressed glass. We propose that the glassy GeO$_2$ adopts therefore a less effective compaction mechanism and a highly metastable structure. Above 158 GPa, the strain energy coupled with the ionizing X-ray beam provides sufficient input energy to crystalize glassy GeO$_2$ into the stable pyrite-like structure.

## 5. Materials and Methods

### 5.1 Synthesis of GeO₂ Glass

Homogeneous and bubble-free $GeO_2$ glass was synthesized at high temperature conditions. In the first step, 2 g of optical-grade $GeO_2$ powder (Sigma-Aldrich, 99.99%) was placed in a Pt crucible and closed with a Pt lid to minimize germanium devolatilization. The crucible was transferred to an electric furnace and heated up to 1600 °C for 30 min, followed by a fast cooling to 1200 °C. The sample was subsequently heated at 1200 °C for 1 h. Then, the crucible was removed from the furnace and immersed immediately in a water bath to quench the melt. The quench rate was estimated to be 400 °C per second and led to bubble-free and homogenous glass formation. The glass was subsequently annealed at 700 ºC for 2 h to remove residual stress. The absence of crystals and air bubbles in the synthetic $GeO_2$ glass was verified using laboratory X-ray diffraction and Raman spectroscopy.

### 5.2 High-Pressure Experiments

High-pressures were generated using membrane-driven diamond anvil cells (m-DAC) with LeToullec design (49). In total, four high-pressure XAS runs were performed covering the pressure range between 0 and 162 GPa, two with and two without pressure transmitting medium (PTM); all details are listed in Table 1. To ensure high-quality XAS data, nano-polycrystalline diamonds were employed (50-52), provided by the University of Ehime in Matsuyama (Japan), with culet sizes between 300 and 100 μm. For all runs, rhenium gaskets were used with a laser-drilled concentric hole in the center of their pre-indentation. For sample loadings without PTM (run 1 and 2), the sample chamber was entirely filled by finely grinded powder of $g$-$GeO_2$ (referred to as solid loading) together with a micrometric ruby sphere for the pressure determination (49). For sample loadings with PTM (run 3 and 4), a polished glass piece with dimensions of $20{\times}20{\times}12$ μm$^3$ was loaded together with Ne as a PTM (53) and a ruby sphere. Beyond 80 GPa the Re gasket (rhenium EOS (54)) was employed as pressure gauges in run 2.

**Table 1.** Conditions of each experimental run. PTM: pressure transmitting medium. *Beveled diamonds.

| Run | $P$ range (GPa) | $P$ gauge | PTM | Diamond's culet size (μm) | Sample thickness (μm) |
|-----|-----------------|-----------|------|---------------------------|------------------------|
| 1 | 0–24 | Ruby | None | 100×300* | 15 |
| 2 | 0–162 | Ruby + Re | None | 150×300* | 15 |
| 3 | 0–40 | Ruby | Ne | 250 | 12 |
| 4 | 0–18 | Ruby | Ne | 300 | 12 |

### 5.3 XAS and XRD Measurements

High-pressure XAS and XRD experiments have been performed at the ESRF beamline BM23 (40). To this end a double-crystal fixed exit monochromator was equipped with two Si(111) crystals and combined with a Pt coated bended Kirkpatrick-Baez mirror system inclined to 4 mrad, used for beam focusing down to $3{\times}3$ μm$^2$ (hor.×ver. beam size at FWHM) and higher harmonic rejection. The monochromator energy was precisely calibrated and verified by acquiring regularly a reference spectrum of a powder pellet of glassy $GeO_2$. The sample was positioned on the μXAS station equipped with submicron precision motorization ideal for repositioning of micrometric samples contained in the diamond anvil cell into the focused beam (31). Ge $K$-edge (11,103 eV) spectra were recorded in transmission geometry using ionization chambers to measure with high precision the beam intensity before and after the sample. Ionization chambers were filled with appropriate gas mixtures for achieving 20% (0.31 Kr + 1.69 He bar) and 80% (1.65 Kr + 0.35 He bar) photon flux absorption, respectively. Ge $K$-edge data were recorded with an energy stepping of 5 eV before the main edge, 0.25 eV at the near-edge range (XANES), and up to a $k$-range of 16 Å$^{-1}$ (EXAFS) with a $k$ stepping of 0.035 Å$^{-1}$. Details on the XANES and EXAFS data analysis procedure can be found in Supporting Information. The μXAS station at BM23 is equipped with a detector block positioned after the samples comprising the second and third ionization chambers, a PRL system (Pressure by Ruby Luminescence), and a Pilatus 1M diffraction detector. More details on the setup can be found in Rosa, *et al.* (40). To verify the amorphous nature of the sample and for pressure determination XRD data were recorded at an incident energy of 18 keV ($\lambda_i$ = 0.6888 Å). The sample-to-detector distance (278.905 mm) and detector tilt parameters were calibrated prior the experiment using a CeO$_2$ standard powder measurements and the software Dioptas (55).

## Author Contributions

The original project was conceived by M.W, A.D.R., J.E.F.S.R., and R.T. The experiments were performed by J.E.F.S.R, A.D.R., E.M., G.G., and O.M. T.I. provided nano-polycrystalline diamonds for the high-pressure X-ray absorption experiments. The data were analyzed and discussed by J.E.F.S.R., A.D.R., and M.W. The manuscript was written by J.E.F.S.R. and A.D.R. with contributions from all the coauthors.

## Notes

The authors declare no competing financial interest.

## Data Availability Statement

The data sets used and/or analyzed are available from the corresponding author on reasonable request.

## Acknowledgements

All the authors wish to express their gratitude to the European Synchrotron Radiation Facility (ESRF, Grenoble) for making all the facilities available for the X-ray absorption spectroscopy measurements in ESRF beamline BM23 (proposal HC-4435). J.E.F.S.R. and AD.R. thank J. Jacobs for his assistance at ESRF's High-Pressure Laboratory and Dr. T. Shinmei for the preparation of the nano-polycrystalline diamonds at the Geodynamics Research Center (Ehime University).

# Absence of higher than 6-fold coordination in glassy GeO$_2$ up to 158 GPa revealed by X-ray absorption spectroscopy

**Authors:**

João Elias F. S. Rodrigues[1,‡], Angelika D. Rosa[1,§], Emin Mijit[1], Tetsuo Irifune[2], G. Garbarino[1], Olivier Mathon[1], Raffaella Torchio[1], Max Wilke[3]

**Affiliations:**

[1] European Synchrotron Radiation Facility (ESRF), 71 Avenue des Martyrs, 38000 Grenoble, France.

[2] Geodynamics Research Center, Ehime University Johoku Campus, 2-5 Bunkyocho, Matsuyama, Ehime 790–0826, Japan.

[3] Institute of Geosciences, University of Potsdam, Karl-Liebknecht-Strasse 24–25, 14476 Potsdam-Golm, Germany.

**Corresponding authors:**

‡ rodrigues.joaoelias@esrf.fr

§ angelika.rosa@esrf.fr

**EXAFS Data Fitting**

XAS data extraction and structural fittings were performed using the Athena and Artemis software packages (1, 2), respectively. The initial data reduction included pre-edge background subtraction, edge jump normalization, and XANES edge position analysis. Based on this reduction, the EXAFS oscillations $\chi(k)$ (the post-edge region) were extracted from the absorption coefficient ($\mu$) and its subtraction from $\mu_0$, in agreement with the equation $\chi(k) + 1 = \mu(k)/\mu_0$, where $\mu_0$ represent the isolated atom absorption coefficient that was found using a polynomial-type background spread function. Then, the extracted and unfiltered EXAFS function $\chi(k)$ was treated using the standard EXAFS equation (3), as given by:

$$\chi(k) = S_0^2 \sum_j N_j \frac{|f_j(k)|}{kR_j^2} \sin[\phi_j(k) + 2kR_j] e^{-2\sigma_j^2 k^2} e^{-2R_j/\lambda(k)} \qquad (1)$$

where, $j$ indexes the single scattering paths, $k = \sqrt{2m_e(E - E_0)/\hbar^2}$ stands for the photoelectron wavenumber in units of Å$^{-1}$, $S_0^2$ the amplitude reduction factor, $N_j$ the number of neighboring atoms or coordination number, $R_j$ the scattering path distance, $\sigma_j^2$ the bond variance or Debye-Waller exponent; $\left|f_j(k)\right|$, $\phi_j(k)$, and $\lambda(k)$ denote the effective backscattering amplitude, its associated phase shift, and photoelectron mean free path, respectively.

For the fitting of high-pressure EXAFS data, the parameter $S_0^2$ was fixed to a value of 0.968. This value was derived from the fit to the spectrum of $g$-GeO$_2$ under room conditions. In this fit, the coordination number of the first shell Ge–O was constrained to be 4 (quartz-like) (4, 5). For all fits, the *FEFF6* code integrated in the Artemis software package was used to determine the *ab initio* functions, including backscattering amplitude, phase shift, and photoelectron mean free path (6). The Fourier transforms (*FT*) of the $k$-weighted EXAFS functions $k^3\chi(k)$ were obtained using a Hanning-type apodization function. To derive the parameters $R_j$, $\sigma_j^2$, and $N_j$ for the high-pressure spectra two methods were used. In the first method, only the next-nearest neighbor Ge–O was considered in the fitting procedure, which included the structural parameters $<R_{Ge–O}>$, $\sigma^2_{R_{Ge–O}}$, and $N_{Ge–O}$ (hereafter, **method 1**). In the second method, two additional paths were included, to account for the second nearest neighbor cation-cation interactions (non-bonded Ge⋯Ge) as well as third nearest neighbors non bonded Ge⋯O interactions (hereafter, **method 2**). For the second method, the parameters $<R_{Ge⋯Ge}>$, $\sigma^2_{R_{Ge⋯Ge}}$, $<R_{Ge⋯O}>$, and $\sigma^2_{R_{Ge⋯O}}$ were fitted, while the number of neighbors $N_j$ were fixed to values found in crystalline structural models that could best describe the data. In particular, $N_{Ge⋯Ge}$ was fixed to 4 for data up to 10 GPa and to 10 beyond 30 GPa, respectively, while $N_{Ge⋯O}$ was fixed to 4 for data up to 10 GPa and to 8 beyond 30 GPa, respectively. The inclusion of the second Ge⋯O path became essential for pressures beyond 9–10 GPa (7).

The fitting method was chosen depending on the raw spectra quality. We have used the first fitting method for run 1 only in which $k$- and $R$-ranges were set to $\Delta k = 2$–8.5 Å$^{-1}$ and $\Delta R = 1.1$–2.2 Å, respectively. For the remaining runs 2–4, the second fitting approach was used with $k$- and $R$-ranges of $\Delta k = 2$–10.5 Å$^{-1}$ and $\Delta R = 1.1$–3.4 Å, respectively. The $\Delta k$ and $\Delta R$ ranges restrict the number of free variables accessed from the fit and, consequently, strong correlations can occur between selected pair of structural parameters, for instance, the well-known correlation of $N_j$ and $\sigma_j^2$ (4, 8). To assess potential correlations, we conducted EXAFS analyses using a **modified method 2** in which the coordination number was held constant at 4, 5, and 6 for XAS data from runs 2–3. This additional approach allowed an accurate tracking of the 4- to 6-fold coordination polyhedron transition by evaluating the fashion of the reliability factors (here, the $R$-factor). Along runs 1 and 4, the edge position was kept fixed to 11116.5 eV. However, along runs 2 and 3, slightly corrections of edge positions were allowed during the fitting through the parameter $\Delta E_0$, which directly affects the photoelectron wavenumber as $k = \sqrt{2m_e(E - E_0 + \Delta E_0)/\hbar^2}$. The variation range of $\Delta E_0$ for each run is listed in Table S1.

The reliability of a structural parameter is highly dependent on the degree of correlation observed between the selected pairs of adjusted variables, typically between $N_j$ and $\sigma_j^2$ or $N_j$ and $S_0^2$ (4, 8). For amorphous GeO$_2$, a successful decorrelation method involves considering the $R$-factor as a relevant parameter to obtain the true coordination number (4, 8). This approach involves keeping $N_{Ge–O}$ fixed at 4 (quartz-like), 5, and 6 (rutile- and CaCl$_2$-like) throughout the fittings. This approach allowed us evaluating which choice of $N_{Ge–O}$ minimizes the $R$-factor. The $R$-factor is plotted as a function of pressure for runs 2–3 in Fig. 2(c-d). For run 2, the quartz-like coordination $N_{Ge–O} = 4$ provides best fit until 10 GPa. For the fits in the range 20–50 GPa, with $N_{Ge-O}$ equal to 5 and 6 similar $R$-factors are obtained. Interestingly, the fit with $N_{Ge-O}$ = 6 consistently minimizes this parameter.

For run 3, the lowest-pressure range 0–7 GPa is well described using 4-fold coordination for first path Ge–O, followed by an intermediate range 7–10 GPa when both $N_{Ge-O}$ of 5 or 6 similarly minimizes $R$-factor. However, the rutile-like environment stabilizes the fit to highest pressure of ~40 GPa, discarding the occurrence of 4- or 5-fold $[GeO_{4,5}]$ units. It is evident that the pressure-transmitting medium is unlikely to play a role in stabilizing 5- or 6-fold coordinated Ge at pressures above ~10 GPa. These tests revealed that the parameter $N_{Ge-O}$ can be set as a free parameter. The obtained $R$-factor values attested the reliability of these fittings when confronted with those values for fixed $N_{Ge-O}$. In Fig. 2(e-h), a comparison is shown between raw and best fitted $k^3\chi(k)$ functions for run 2 at selected pressures. This elucidates the achievement of a reasonable EXAFS model for the high-pressure local structure of glassy $GeO_2$.

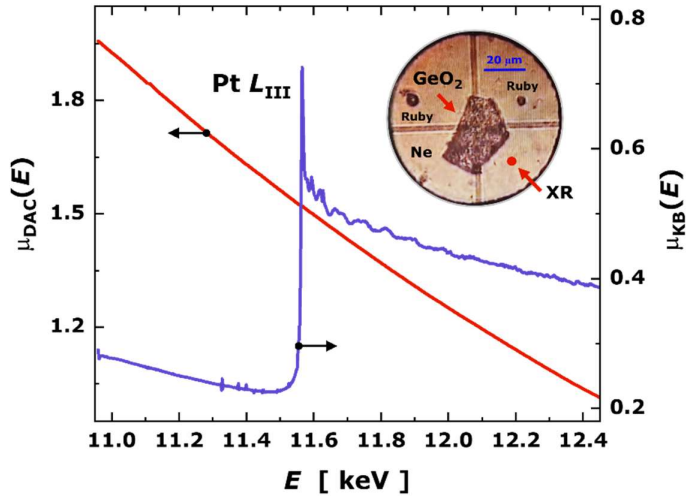

**Figure S1.** Red line: X-ray absorption spectrum taken through the diamond anvil cell on the neon pressure medium at ~34 GPa. The circular inlet shows a photograph of the sample recorded through one diamond and the position of the X-ray beam for the measurement (red dot). The rectangular polished $GeO_2$ glass piece is placed in the center of the Re gasket hole (black rim) together with two ruby spheres and neon as pressure transmitting medium. Note that continuous and regular decrease of the signal, which indicates a low background noise and high-quality of the signal normalization. **Blue line:** X-ray absorption spectrum taken using ionization chambers placed before and after the Kirkpatrick-Baez focusing system. Note that in this spectrum the signal coming from Pt coating (Pt $L_{III}$-edge) of the Kirkpatrick-Baez bent mirrors is present. In the red spectrum this noise signal is effectively filtered out, indicating a very high signal to noise ratio for data taken in the DAC.

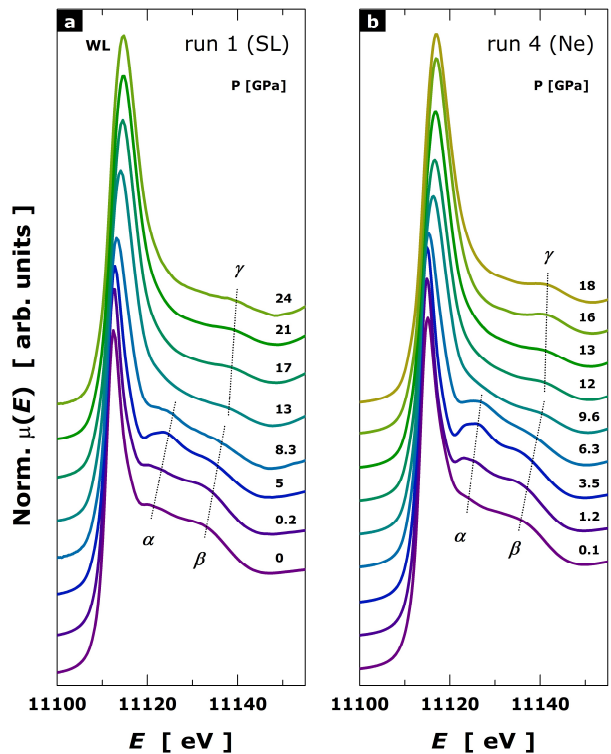

**Figure S2.** Normalized Ge *K*-edge XANES spectra as a function of pressure in the ranges 0–24 GPa (run 1) (a) and 0.1–18 GPa (run 4) (b). The Greek letters $\alpha$, $\beta$, and $\gamma$ represent the spectral features at low- and high-pressure, respectively.

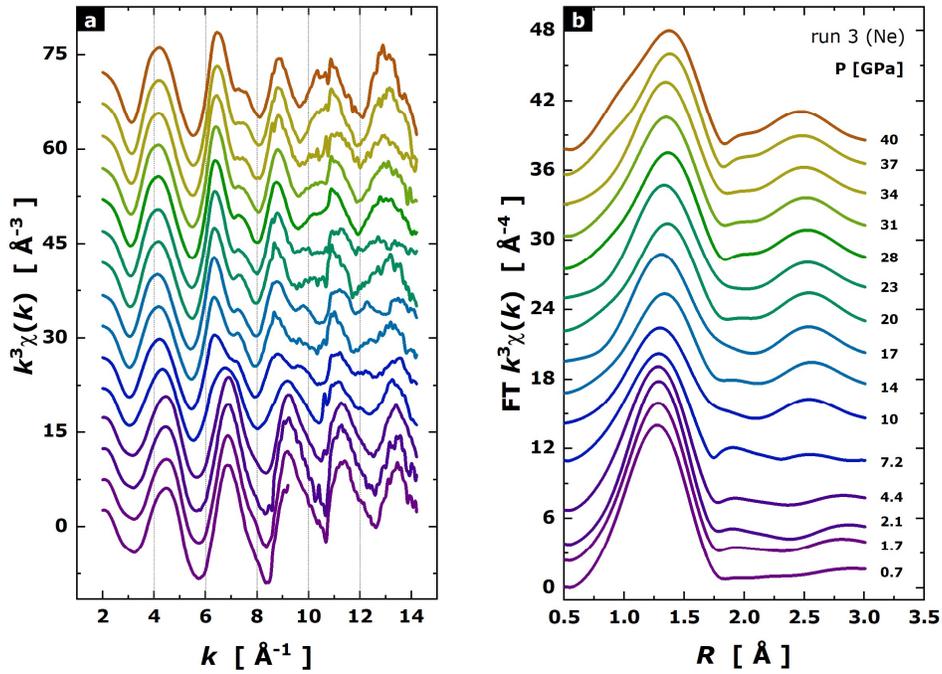

**Figure S3.** Pressure evolution of $k^3$-weighted EXAFS functions $\chi(k)$ (a) and the respective magnitudes of the Fourier transformations (b) in the pressure range 0.7–40 GPa for run 3. The Fourier transformations magnitudes are not phase-shift corrected.

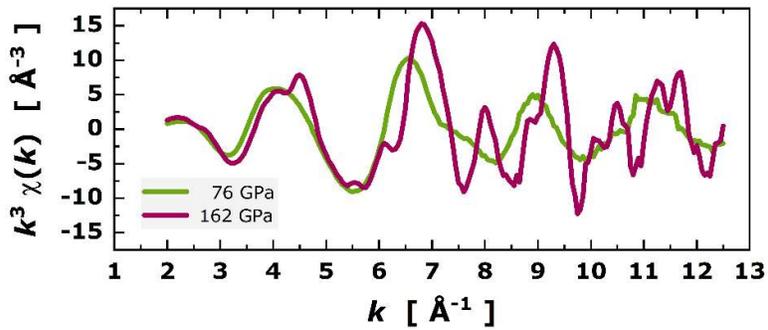

**Figure S4.** Comparison between raw $k^3$-weighted EXAFS functions of glassy $GeO_2$ at 76 GPa and at crystalline $GeO_2$ at 162 GPa. The emergence of sharp high-amplitude oscillations in the EXAFS functions evidence of the crystalline nature of the sample which has been also verified through X-ray diffraction measurements (see Fig. S5).

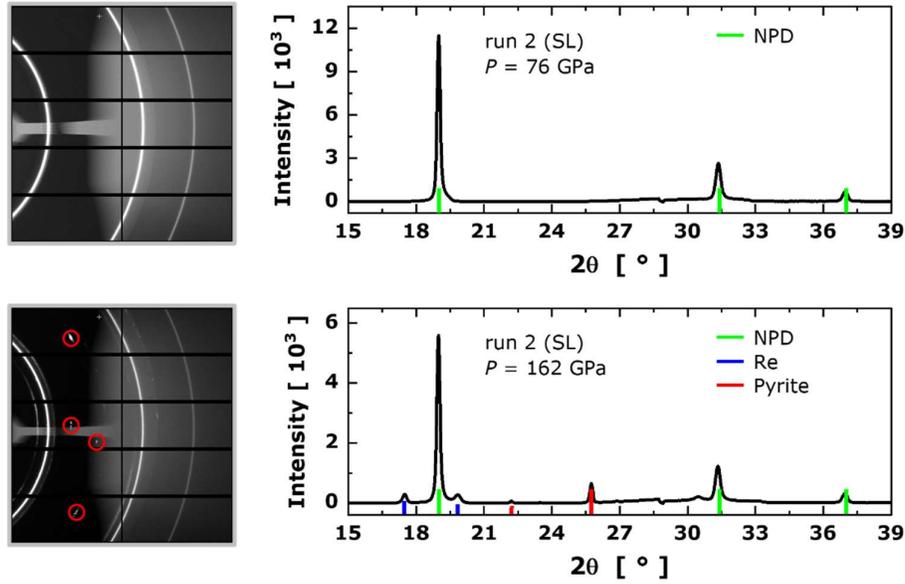

**Figure S5. Top panel:** X-ray diffraction pattern of glassy GeO₂ at 76 GPa from run 2. Note, the absence of crystalline peaks, except those originating from nano-polycrystalline diamonds. **Bottom panel:** Diffraction peaks from the pyrite-like phase of GeO₂ emerged at 159 and 162 GPa, like those reported by Shiraki et al. To the left of each X-ray pattern, the corresponding 2D image is presented. Red circles denote crystalline peaks from the pyrite phase.

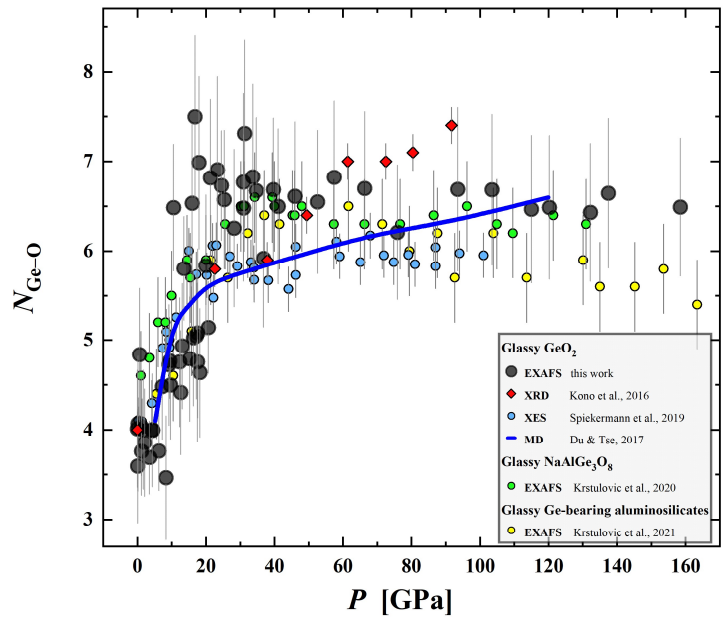

**Figure S6.** Pressure evolution of the Ge coordination number in glassy GeO₂ extracted from EXAFS in this work ($N_{Ge-O}$, gray circles) and those reported by Kono et al from XRD (red diamonds), Spiekermann et al from XES (blue circle), and Du and Tse from MD simulations (blue line). For comparison, the pressure evolution of the Ge coordination number derived from EXAFS in aluminogermanate [Krstulovic et al (2020), green circles] and aluminosilicate glasses are shown [Krstulovic et al (2021), yellow circles]. Pressure uncertainties are smaller than the symbol size.

**Table S1.** Summary of the EXAFS fitting results for the next nearest neighbor Ge–O scattering path ($<R_{Ge-O}>$, $\sigma^2_{RGe-O}$, and $N_{Ge-O}$) and the second next nearest neighbor contribution Ge···Ge distance $<R_{Ge···Ge}>$ for all runs with their uncertainties. The $R$-factors are listed for each fit and are a measure of the goodness of the fit. for each run the pressure and its uncertainty are given. Abbreviations are $P_R$: Ruby pressure and $P_{Re}$: Rhenium pressure.

| Run | $P_R$ (GPa) | $P_{Re}$ (GPa) | $R_{Ge-O}$ (Å) | $\sigma^2_{Ge-O}$ (Å²) | $N_{Ge-O}$ | $R_{Ge···Ge}$ (Å) | $R$-factor |
|---|---|---|---|---|---|---|---|
| **1** | 0 | | 1.739(6) | 0.0012(9) | 4.0(7) | | 0.0194 |
| | 0.2(0) | | 1.725(7) | 0.0013(10) | 4.1(8) | | 0.0229 |
| | 8.3(2) | | 1.797(9) | 0.0024(15) | 3.5(7) | | 0.0138 |
| | 12.7(7) | | 1.833(7) | 0.0066(26) | 4.4(7) | | 0.0358 |
| | 17.6(1) | | 1.854(9) | 0.0075(32) | 4.8(9) | | 0.0547 |
| | 20.7(2) | | 1.851(10) | 0.0077(34) | 5.1(5) | | 0.0611 |
| | 24.5(1) | | 1.837(8) | 0.0089(13) | 6.7(6) | | 0.0511 |
| | | | | | | | |
| | $\Delta k$ (Å⁻¹) | 2–8.5 | $E_0$ (eV) | 11116.5 | | | |
| | $\Delta R$ (Å) | 1.1–2.2 | $|\Delta E_0|$ (eV) | 0 | | | |
| | | | | | | | |
| **2** | 4.4(2) | | 1.735(9) | 0.0010(8) | 4.0(0) | | 0.0181 |
| | 9.4(4) | | 1.748(5) | 0.0044(13) | 4.8(4) | | 0.0043 |
| | 16.0(16) | | 1.807(9) | 0.0104(23) | 6.5(9) | 2.88(3) | 0.0218 |
| | 18.0(4) | | 1.811(9) | 0.0108(23) | 7.0(10) | 2.88(3) | 0.0197 |
| | 21.3(4) | | 1.823(8) | 0.0103(22) | 6.8(9) | 2.87(3) | 0.0208 |
| | 25.4(4) | | 1.831(8) | 0.0091(20) | 6.6(8) | 2.86(2) | 0.0199 |
| | 31.0(0) | | 1.830(8) | 0.0084(21) | 6.5(8) | 2.85(3) | 0.0207 |
| | 31.3(3) | | 1.826(9) | 0.0094(24) | 7.3(10) | 2.85(3) | 0.0237 |
| | 34.7(3) | | 1.830(8) | 0.0084(20) | 6.7(8) | 2.85(3) | 0.0347 |
| | 41.1(6) | | 1.828(8) | 0.0077(22) | 6.5(9) | 2.83(3) | 0.0201 |
| | 46.0(10) | | 1.826(8) | 0.0077(20) | 6.6(8) | 2.83(3) | 0.0211 |
| | 52.6(4) | | 1.827(8) | 0.0076(20) | 6.5(8) | 2.82(2) | 0.0186 |
| | 57.4(3) | | 1.821(8) | 0.0076(20) | 6.8(8) | 2.82(3) | 0.0193 |
| | 66.3(1) | | 1.816(8) | 0.0073(20) | 6.7(9) | 2.81(2) | 0.0183 |
| | 75.9(9) | | 1.804(7) | 0.0062(18) | 6.2(8) | 2.79(2) | 0.0154 |
| | | 93.5(37) | 1.788(8) | 0.0062(20) | 6.7(9) | 2.77(3) | 0.0181 |
| | | 103.5(0) | 1.786(8) | 0.0066(19) | 6.7(8) | 2.78(3) | 0.0173 |
| | | 115.0(40) | 1.783(8) | 0.0064(19) | 6.5(8) | 2.77(2) | 0.0147 |
| | | 120.3(48) | 1.783(7) | 0.0065(19) | 6.5(8) | 2.77(2) | 0.0139 |
| | | 132.3(48) | 1.777(7) | 0.0063(18) | 6.4(8) | 2.76(3) | 0.0137 |
| | | 137.5(20) | 1.771(7) | 0.0063(18) | 6.6(8) | 2.75(2) | 0.0141 |
| | | 158.5(15) | 1.772(7) | 0.0063(17) | 6.5(8) | 2.75(2) | 0.0162 |
| | | | | | | | |
| | $\Delta k$ (Å⁻¹) | 2–10.5 | $E_0$ (eV) | 11116.5 | | | |
| | $\Delta R$ (Å) | 1.1–3.4 | $|\Delta E_0|$ (eV) | 0–2.5 | | | |
| | | | | | | | |
| **3** | 0.7(0) | | 1.740(7) | 0.0032(10) | 4.8(7) | 3.13(2) | 0.0127 |
| | 1.7(1) | | 1.736(7) | 0.0022(10) | 4.0(7) | 3.12(2) | 0.0132 |
| | 2.1(0) | | 1.738(6) | 0.0016(9) | 3.9(6) | 3.11(3) | 0.0114 |
| | 4.4(2) | | 1.740(7) | 0.0029(9) | 4.0(6) | 3.10(3) | 0.0120 |
| | 7.2(3) | | 1.754(6) | 0.0063(9) | 4.5(6) | 2.84(5) | 0.0081 |
| | 10.5(6) | | 1.796(13) | 0.0107(20) | 6.5(7) | 2.87(3) | 0.0176 |
| | 13.6(4) | | 1.812(11) | 0.0093(18) | 5.8(6) | 2.87(3) | 0.0158 |
| | 16.8(2) | | 1.814(15) | 0.0104(23) | 7.5(9) | 2.85(4) | 0.0232 |
| | 20.0(4) | | 1.826(9) | 0.0085(15) | 5.8(8) | 2.86(2) | 0.0114 |
| | 23.4(6) | | 1.820(11) | 0.0090(17) | 6.9(10) | 2.84(3) | 0.0144 |
| | 28.2(3) | | 1.830(10) | 0.0083(16) | 6.3(9) | 2.89(3) | 0.0129 |
| | 30.9(2) | | 1.822(11) | 0.0085(16) | 6.8(10) | 2.87(3) | 0.0135 |
| | 33.7(2) | | 1.823(11) | 0.0083(17) | 6.8(10) | 2.87(3) | 0.0149 |
| | 36.8(2) | | 1.829(9) | 0.0076(14) | 5.9(8) | 2.87(3) | 0.0116 |
| | 39.7(4) | | 1.825(9) | 0.0086(14) | 6.7(8) | 2.87(3) | 0.0097 |
| | | | | | | | |
| | $\Delta k$ (Å$^{-1}$) | 2–10.5 | $E_0$ (eV) | 11116.5 | | | |
| | $\Delta R$ (Å) | 1.1–3.4 | $|\Delta E_0|$ (eV) | 0–3.0 | | | |
| | | | | | | | |
| **4** | 0.1(1) | | 1.755(11) | 0.0009(3) | 3.6(6) | 3.16(4) | 0.0130 |
| | 1.2(0) | | 1.744(5) | 0.0007(3) | 3.8(5) | 3.15(2) | 0.0238 |
| | 1.2(0) | | 1.748(6) | 0.0019(5) | 4.0(5) | 3.15(2) | 0.0260 |
| | 3.5(2) | | 1.750(5) | 0.0011(9) | 3.7(4) | 3.12(1) | 0.0174 |
| | 6.3(4) | | 1.762(6) | 0.0038(13) | 3.8(4) | 3.07(4) | 0.0199 |
| | 9.2(1.1) | | 1.838(10) | 0.0054(24) | 4.7(7) | 2.88(4) | 0.0130 |
| | 9.6(3) | | 1.812(8) | 0.0071(17) | 4.5(6) | 2.89(3) | 0.0191 |
| | 12.3(2) | | 1.820(9) | 0.0070(20) | 4.8(7) | 2.88(3) | 0.0245 |
| | 13.1(9) | | 1.841(11) | 0.0063(25) | 4.9(7) | 2.88(3) | 0.0121 |
| | 15.2(8) | | 1.826(9) | 0.0068(19) | 4.8(7) | 2.87(2) | 0.0225 |
| | 16.4(1) | | 1.836(8) | 0.0050(20) | 5.0(8) | 2.84(2) | 0.0124 |
| | 17.2(10) | | 1.839(9) | 0.0078(19) | 5.0(7) | 2.89(2) | 0.0212 |
| | 17.6(3) | | 1.832(10) | 0.0057(23) | 5.1(9) | 2.86(3) | 0.0259 |
| | 18.2(0) | | 1.821(9) | 0.0053(18) | 4.6(7) | 2.86(2) | 0.0271 |
| | | | | | | | |
| | $\Delta k$ (Å$^{-1}$) | 2–10.5 | $E_0$ (eV) | 11116.5 | | | |
| | $\Delta R$ (Å) | 1.1–3.4 | $|\Delta E_0|$ (eV) | 0 | | | |

**Bond valence theory**

In the bond-valance sum theory developed by Brown (9), the "cation bond strength" ($s$) is related to the ratio between the cations' charge ($Z_C$) and the average coordination number $<Nc>$, it represents the valence for a single bond formed by the cation and oxygen. Within this theory, $s$ can be also empirically related to the difference between the experimentally determined cation-anion distance ($<R_{Ge-O}>$) and an ideal cation-anion bond length $R_0$ with unit valance. Therefore, $Nc$ and $R$ parameters are inversely correlated through the relation:

$$s = \frac{Z_C}{Nc} = \frac{R - R_c}{e^{0.37}} \qquad (2)$$

We assume that the cation charge should be constant throughout the investigated pressure range, which implies that the sum of the bond strength values is constant. Finally, we consider that the structural arrangement of cations and anions is such that local charge neutrality is achieved. The bond length $R_0$ which represents the bond length for $s = 1$ is determined from known crystalline structures and must be adjusted for high pressure to account for the changes in the balance between repulsive and attractive forces as, discussed in Brown et al (9, 10).

**The cation-cation repulsion theory**

According to O'Keefe and Hyde (11), the coordination number ($CN$) of a system is the result of the balance of attractive and repulsive interactions between bonded cations and anions and the repulsive forces between non-bonded cations. This theory also considers that the coordination numbers of both the anion ($CN_a$) and the cation ($CN_c$) are correlated to the stoichiometric cation-anion ratio. For example, in the case of $GeO_2$, the cation-anion ratio is always 0.5 which implies a ratio 1:2 for $CN_a : CN_c$ [for the quartz(rutile) structure of $GeO_2$]. Ge is four(six)-fold coordinated, while O is two(three)-fold coordinated in the quartz-type structure). Any increase of $CN_c$ implies therefore an increase of $CN_a$ and vice versa.

The controlling parameter for $CN_a$ is the ratio between the non-bonded cation radius ($R_{Ge}$) determined from $<R_{Ge\cdots Ge}>/2 = (R_{Ge}) = 1.576$ Å for Ge in α-quartz), and the bonded cation-anion bond length ($l = <R_{Ge-O}>$). The ratio $R_{Ge}/l$ is distinct for specific anion coordination numbers, $i.e.$, $CN_a = 2$ requires that two cations are opposed to each other, therefore: $R_{Ge}/l \le \sin(180°/2) \le 1$. If the ratio $R_{Ge}/l$ decreases, $CN_a$ increases: if $CN_a = 3$ the ratio $R_{Ge}/l$ needs to be equal or smaller than $\sin(180°/3) \le 0.866$ to fit 3 cations around an anion. It follows that for $CN_a = 4$ the ratio $R_{Ge}/l \le \sin(109.28°/2) \le 0.8160$ and for $CN_a = 6$ the ratio $R_{Ge}/l \le \sin(90°/2) \le 0.707$ while for $CN_a = 8$ it follows that $R_{Ge}/l = 0.577$. Any coordination number increase with pressure can be, therefore, understood as the reduction of $<R_{Ge\cdots Ge}>$ non-bonded distances, in which pressure overcomes the repulsive forces between cations that are arranged around an anion. In <span style="color:blue">Fig. S7</span>, the pressure evolution of $R_{Ge}/l$ is plotted for runs 2–3 of glassy $GeO_2$. The distances $<R_{Ge-O}>$ and $<R_{Ge\cdots Ge}>$ were estimated from EXAFS, as listed in <span style="color:blue">Table S1</span>. At ambient conditions, the $R_{Ge}/l$ ratio in α-quartz $GeO_2$ is 0.91 (1.576 Å / 1.737 Å), which confirms a $CN_a$ of 2. The geometric concept outlined above holds for corner-sharing polyhedra. For edge-sharing polyhedra, such as in the rutile-structure polymorph, a two-angle situation must be considered. The edge-sharing polyhedra induce higher repulsive cation-cation interactions and thus increase repulsive energy. This energy is compensated with the formation of additional anion-cation bonds. The core principle of this theory is, therefore, that any increase in anion coordination follows a reduction of $<R_{Ge\cdots Ge}>$ if polyhedral connectivity remains unchanged.

This principle is qualitatively used to interpret the pressure evolutions of the extracted bonded anion-cation and non-bonded cation-cation distances in terms of a pressure-induced coordination change in glassy GeO₂, based on the data represented in Fig. S7. As can been seen from the pressure evolution of the $R/l$ ration shown in Fig. S7, a clear reduction of the $R/l$ ratio can be observed between 10–30 GPa, the interval in which the 4-to-6-fold transition takes place and corner sharing tetrahedra evolves to edge and corner sharing octahedral. Interestingly the $R/l$ ratio does not further evolve with pressure suggesting that the structural arrangement remains.

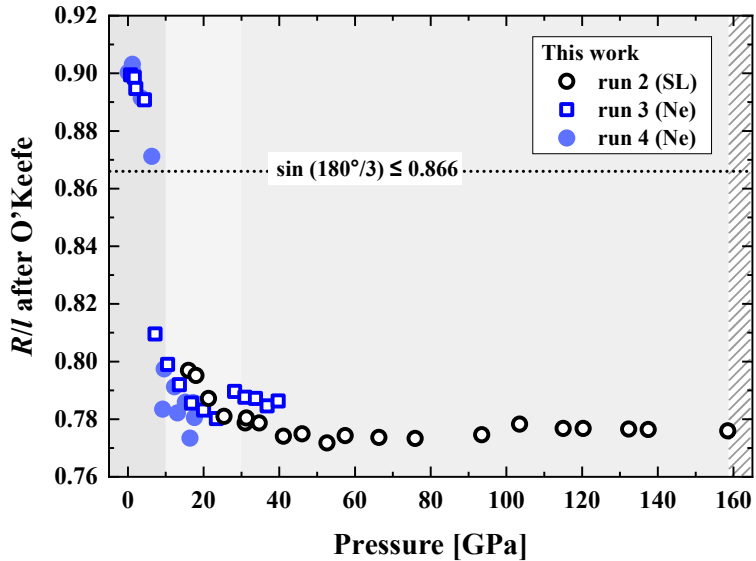

**Figure S7.** Pressure evolution of the $R_{Ge}/l$ ratio in $g$-GeO₂ extracted from distances $<R_{Ge-O}>$ and $<R_{Ge-Ge}>$ (summarized in Table S1).